\documentclass[12pt]{article}
\usepackage{amsmath,color}
\usepackage[dvips]{graphicx}
\input amssym
\input amssym.def

\def\red#1{{\color{red} #1}}
\begin{document}

\def\prg#1{\medskip\noindent{\bf #1}}   \def\ra{\rightarrow}
\def\lra{\leftrightarrow}               \def\Ra{\Rightarrow}
\def\nin{\noindent}                     \def\pd{\partial}
\def\dis{\displaystyle}                 \def\Lra{{\Leftrightarrow}}
\def\cs{{\scriptstyle\rm CS}}           \def\ads3{{\rm AdS$_3$}}
\def\Leff{\hbox{$\mit\L_{\hspace{.6pt}\rm eff}\,$}}
\def\bull{\raise.25ex\hbox{\vrule height.8ex width.8ex}}
\def\bul{\raise.25ex\hbox{\vrule height.4ex width.4ex}}
\def\ric{{Ric}}                         \def\tric{{(\widetilde{Ric})}}
\def\Lie{{\cal L}\hspace{-.7em}\raise.25ex\hbox{--}\hspace{.2em}}
\def\hd{{^\star}}                       \def\ul#1{\underline{#1}}
\def\mb#1{\hbox{{\boldmath $#1$}}}      \def\grp{{GR$_\parallel$}}
\def\irr#1{^{(#1)}}                     \def\Ld{\cL}
\def\tgr{GR$_\parallel$}

\def\hook{\hbox{\vrule height0pt width4pt depth0.3pt
\vrule height7pt width0.3pt depth0.3pt
\vrule height0pt width2pt depth0pt}\hspace{0.8pt}}
\def\semidirect{\;{\rlap{$\supset$}\times}\;}
\def\inn{\hook}
\def\bm#1{\hbox{{\boldmath $#1$}}}
\def\nb#1{\marginpar{{\large\bf #1}}}
\def\ir#1{{}^{(#1)}}  \def\orth{{\perp}}

\def\G{\Gamma}        \def\S{\Sigma}        \def\L{{\mit\Lambda}}
\def\D{\Delta}        \def\Th{\Theta}       \def\cP{{\cal P}}
\def\a{\alpha}        \def\b{\beta}         \def\g{\gamma}
\def\d{\delta}        \def\m{\mu}           \def\n{\nu}
\def\th{\theta}       \def\k{\kappa}        \def\l{\lambda}
\def\vphi{\varphi}    \def\ve{\varepsilon}  \def\p{\pi}
\def\r{\rho}          \def\Om{\Omega}       \def\om{\omega}
\def\s{\sigma}        \def\t{\tau}          \def\eps{\epsilon}
\def\nab{\nabla}      \def\btz{{\rm BTZ}}   \def\bR{{\bar R}}
\def\heps{\hat\eps}   \def\tD{{\tilde\nab}} \def\bu{{\bar u}}
\def\bv{{\bar v}}     \def\bs{{\bar s}}     \def\bx{{\bar x}}
\def\by{{\bar y}}     \def\bom{{\bar\om}}   \def\tphi{{\tilde\vphi}}  \def\tt{{\tilde t}}   \def\tcV{{\tilde{\cal V}}} \def\tV{{\tilde V}}
\def\hpi{{\hat\pi}}   \def\bm{{\bar m}}     \def\bn{{\bar n}}
\def\bi{{\bar\imath}} \def\bj{{\bar\jmath}} \def\bk{{\bar k}}

\def\tG{{\tilde G}}   \def\cF{{\cal F}}      \def\bH{{\bar H}}
\def\cL{{\cal L}}     \def\bcL{{\bar\cL}}    \def\hL{{\hat\L}}
\def\tcL{{\tilde\cL}} \def\cM{{\cal M }}     \def\cE{{\cal E}}
\def\cV{{\cal V}}     \def\hcV{{\hat\cV}}
\def\cH{{\cal H}}     \def\hcH{\hat{\cH}}    \def\hH{\hat{H}}
\def\cA{{\cal A}}     \def\cT{{\cal T}}      \def\cB{{\cal B}}
\def\cK{{\cal K}}     \def\hcK{\hat{\cK}}    \def\cE{{\cal E}}
\def\cR{{\cal R}}     \def\hcR{{\hat\cR}}    \def\hR{{\hat R}{}}
\def\cO{{\cal O}}     \def\hcO{\hat{\cal O}} \def\tom{{\tilde\om}}
\def\tA{{\tilde A}}   \def\tT{{\tilde T}}    \def\tR{{\tilde R}}
\def\hy{{\hat y}\hspace{1pt}}  \def\tcO{{\tilde\cO}}
\def\bA{{\bar A}}     \def\bB{{\bar B}}      \def\bC{{\bar C}}
\def\bG{{\bar G}}     \def\bD{{\bar D}}      \def\bH{{\bar H}}
\def\bK{{\bar K}}     \def\bL{{\bar L}}      \def\hcT{{\hat\cT}}
\def\hcR{{\hat\cR}}   \let\Pi\varPi          \def\bk{{\bar k}}

\def\rdc#1{\hfill\hbox{{\small\texttt{#1}}}}
\def\chm{\checkmark}  \def\chmr{\red{\chm}}
\def\dual#1{\hspace{0.5pt}{}^{(\star)}\hspace{-.3pt}#1}
\def\nn{\nonumber}                    \def\vsm{\vspace{-9pt}}
\def\be{\begin{equation}}             \def\ee{\end{equation}}
\def\ba#1{\begin{array}{#1}}          \def\ea{\end{array}}
\def\bea{\begin{eqnarray} }           \def\eea{\end{eqnarray} }
\def\beann{\begin{eqnarray*} }        \def\eeann{\end{eqnarray*} }
\def\beal{\begin{eqalign}}            \def\eeal{\end{eqalign}}
\def\lab#1{\label{eq:#1}}             \def\eq#1{(\ref{eq:#1})}
\def\bsubeq{\begin{subequations}}     \def\esubeq{\end{subequations}}
\def\bitem{\begin{itemize}}           \def\eitem{\end{itemize}}
\renewcommand{\theequation}{\thesection.\arabic{equation}}
\title{Entropy in Poincar\'e gauge theory:\\ Hamiltonian approach}

\author{M. Blagojevi\'c and B. Cvetkovi\'c\footnote{
        Email addresses: \text{mb@ipb.ac.rs, cbranislav@ipb.ac.rs}} \\
Institute of Physics, University of Belgrade,\\
                      Pregrevica 118, 11080 Belgrade-Zemun, Serbia}
\date{\today}
\maketitle

\begin{abstract}
The canonical generator $G$ of local symmetries in Poincar\'e gauge theory is constructed as an integral over a spatial section $\S$ of spacetime. Its regularity (differentiability) on the phase space is ensured by adding a suitable surface term, an integral over the boundary of $\S$ at infinity,  which represents the asymptotic canonical charge. For black hole solutions, $\S$ has two boundaries, one at infinity and the other at horizon. It is shown that the canonical charge at horizon defines entropy, whereas the  regularity of $G$ implies the first law of black hole thermodynamics.
\end{abstract}

\section{Introduction}   
\setcounter{equation}{0}

In the early 1960s, Kibble and Sciama \cite{x1} proposed a new theory of gravity, the Poincar\'e gauge theory (PG), based on gauging the Poincar\'e group of spacetime symmetries. By construction, PG is characterized by a Riemann-Cartan (RC) \emph{geometry} of spacetime, in which both the torsion and the curvature are essential ingredients of the \emph{gravitational dynamics}. Nowadays, PG is a well-established approach to gravity, representing a natural gauge-field-theoretic extension of general relativity (GR). For more details, see review articles by Hehl et al. \cite{x2}, a reader with commentaries by Blagojevi\'c and Hehl \cite{x3}, and monographs by Blagojevi\'c \cite{x4}, Ponomariov et al. \cite{x5}, and Mielke \cite{x6}.

In the past half century, many investigations of PG have been aimed at clarifying different aspects of both the geometric and dynamical roles of torsion. In particular, successes in constructing exact solutions with torsion naturally raised the question of how their \emph{conserved charges} are influenced by the presence of torsion; for a review, see Ref. \cite{x3}. Relying on these developments, we will reconsider the notion of conserved charge in the Hamiltonian formalism, as it represents the most natural basis for the main subject of the present paper, the \emph{influence of torsion} on black hole entropy.

The expressions for the conserved charges in PG were first found for asymptotically flat solutions \cite{x7,x8}. The results obtained by Blagojevi\'c and Vasili\'c \cite{x8} are based on the Hamiltonian approach to PG \cite{x9,x10} combined with the ideas of Regge and Teitelboim \cite{x11}. In this approach, the conserved charges are represented by a boundary term, defined by requiring the variation of the canonical gauge generator to be a well-defined (differentiable) functional on the phase space. A covariant version of the Hamiltonian approach, introduced later by Nester \cite{x12}, turned out to be an important step in understanding the conservation laws. This was clearly demonstrated by Hecht and Nester \cite{x13,x14}, in their analysis of the conserved charges for asymptotically flat or (anti) de Sitter solutions. Further development of these ideas can be found in Nester and co-workers \cite{x15}, and a comprehensive exposition, incorporating the latest developments, is given by Chen et al. \cite{x16}.

Despite such an intensive activity in exploring the notion of conserved charges in the \emph{generic} four-dimensional (4D) PG,  systematic studies of black hole entropy in the presence of torsion have been largely neglected in the literature. One should mention here an early and rather general proposal by Nester and co-workers \cite{x15} which, however, did not prove to be quite successful. Investigations of black hole entropy in Refs. \cite{x17} are restricted to a specific version of PG, the Einstein-Cartan theory, which is certainly not sufficient to justify any conclusion on the general relation between torsion and entropy. In 3D  gravity, black hole entropy is well understood for solutions possessing the asymptotic conformal symmetry \cite{x18,x19,x20}, but for 4D black holes, such an approach is much less settled \cite{x21}.

The physics of black holes is an arena where thermodynamics, gravity, and
quantum theory are connected through the existence of entropy as an intrinsic dynamical aspect of black holes \cite{x22}. In the period around the 1990s, understanding of the \emph{classical} black hole entropy reached a level that can be best characterized by Wald's words: ``Black hole entropy is the Noether charge" \cite{x23}. The question that we wish to address in the present paper is  whether such a challenging idea, transformed to a more natural canonical framework, can improve our understanding of black hole entropy in the generic PG.

The paper is organized as follows. In section 2, we give a short account of the Lagrangian formalism for PG, and discuss the notion of surface gravity.
In section 3, we develop the canonical formulation of PG in the first order formalism and use it to construct the canonical gauge generator. In section 4, we use the improved form of the gauge generator to obtain the variational equation for the asymptotic canonical charge (energy and angular momentum), located at the spatial 2-boundary at infinity. Then, following the idea that ``entropy is the canonical charge at horizon," we are naturally led to define black hole entropy by the same variational equation, but located at black hole horizon. It is shown that the condition of differentiability of the gauge generator guarantees the validity of the first law of black hole thermodynamics. In sections 5 and 6, our results are tested on three illustrative examples, belonging to the family of spherically symmetric solutions of PG. Finally, section 7 is devoted to concluding remarks, and the appendixes contain important technical details.

Our conventions are as follows.
The greek indices $(\m,\n,\dots)$ refer to the coordinate frame, with a time-space splitting expressed by $\m=(0,\a)$, the latin indices $(i,j,\dots)$ refer to the local Lorentz frame, $b^i$ is the orthonormal tetrad (1-form), $h_i$ is the dual basis (frame), with $h_i\inn b^k = \d_i^k$, and the Lorentz metric is $\eta_{ij}=(1,-1,-1,-1)$. The volume 4-form is $\heps = b^0∧ b^1∧ b^2∧b^3$, the Hodge dual of a form $\a$ is $\hd\a$, with $\hd 1=\heps$, and the totally antisymmetric tensor is defined by $\hd(b_ib_jb_mb_n)=\ve_{ijmn}$, where $\ve_{0123}=+1$. The exterior product of forms is implicit, except in Appendix \ref{appA}.

\section{Preliminaries: PG dynamics and surface gravity} 
\setcounter{equation}{0}

As a preparation for discussing the notion of black hole entropy in PG, it is necessary to clarify to what extent is the existence of torsion compatible with the standard interpretation of surface gravity, introduced in GR.

\subsection{A brief account of PG}

Basic dynamical variables of PG are the tetrad field $b^i$ and the spin connection $\om^{ij}$ (1-forms), the gauge potentials related to the translation and the Lorentz subgroups of the Poincar\'e group, respectively. The corresponding field strengths are the torsion $T^i=db^i+\om^i{_m}b^m$ and the curvature $R^{ij}=d\om^{ij}+\om^i{_m}\om^{mj}$ (2-forms), and the underlying spacetime continuum is characterized by a RC geometry, see for instance \cite{x3,x4,x24}.

Varying the gravitational Lagrangian $L_G=L_G(b^i,T^i,R^{ij})$ (4-form) with respect to $b^i$ and $\om^{ij}$ yields the gravitational field equations \emph{in vacuum}. After introducing the covariant field momenta,
$H_i:=\pd L_G/\pd T^i$ and $H_{ij}:=\pd L_G/\pd R^{ij}$, and the associated energy-momentum and spin currents, $E_i:=\pd L_G/\pd b^i$ and
$E_{ij}:=\pd L_G/\pd\om^{ij}$, these equations can be written in a compact form as
\bsubeq\lab{2.1}
\bea
\d b^i:&&\nab H_i+E_i=0\,,                                      \lab{2.1a}\\
\d\om^{ij}:&&\nab H_{ij}+E_{ij}=0\,.                            \lab{2.1b}
\eea
\esubeq
Explicit expressions for the gravitational currents read
\bea
&&E_i=h_i\inn L_G-(h_i\inn T^m)H_m-\frac{1}{2}(h_i\inn R^{mn})H_{mn}\,,\nn\\
&&E_{ij}=-(b_i H_j-b_j H_i)\, .
\eea

Assuming the gravitational Lagrangian $L_G$ to be at most quadratic in the field strengths (quadratic PG) and parity invariant,
\be
L_G=-\hd(a_0R+2\L)+T^i\sum_{n=1}^3\hd(a_n\ir{n}T_i)
            +\frac{1}{2}R^{ij}\sum_{n=1}^6\hd(b_n\ir{n}R_{ij})\,,  \lab{2.3}
\ee
the gravitational field momenta take the form
\bsubeq\lab{2.4}
\bea
&&H_i=2\sum_{m=1}^3\hd({a_m}\,{}^{(m)}T_i)\, ,                       \\
&&H_{ij}=-2a_0\hd(b^ib^j)+H'_{ij}\, ,\qquad
  H'_{ij}:=2\sum_{n=1}^6\hd({b_n}\,{}^{(n)}R_{ij})\, .
\eea
\esubeq
Here, $(a_0,a_m,b_n)$ are the Lagrangian parameters, $a_0$ is normalized to $16\pi a_0=1$ (in units $G=1$), $\L$ is a cosmological constant,  and $\ir{m}T_i$ and $\ir{n}R_{ij}$ are irreducible parts of the torsion and the curvature, see Appendix \ref{appA}.

In the presence of matter, the right-hand sides of \eq{2.1a} and \eq{2.1b} contain the corresponding matter currents.

\subsection{Surface gravity}

A black hole can be described as a region of spacetime which is causally disconnected from the rest of spacetime. The causal structure of spacetime is most naturally characterized with the help of null geodesics. The boundary of a black hole is a null hypersurface, known as the \emph{event horizon}.

To introduce surface gravity, consider a black hole characterized by the existence of a Killing vector field $\xi$. Then, a null hypersurface to which the Killing vector is normal, is called the \emph{Killing horizon} ($\cK$). As a consequence, $\xi^2:=g_{\m\n}\xi^\m\xi^\n=0$ on $\cK$. Then, since the gradient $\pd_\m(\xi^2)$ is also normal to $\cK$, it must be proportional to $\xi_\m$,
\be
\pd_\m(\xi^2)=-2\k\xi_\m\, ,                                       \lab{2.5}
\ee
where the scalar function $\k$ is known as \emph{surface gravity} \cite{x22,x25}. To have a physical interpretation of surface gravity, the event horizon of a black is must be a Killing horizon. One can show, without making use of any field equations, that indeed, for a wide class of stationary black holes (systems in ``equilibrium"), the Killing horizon coincides with event horizon.

The essential property of surface gravity is expressed by the \emph{zeroth law} of black hole mechanics: For a wide class of stationary black holes, surface gravity is constant over the entire event horizon. For $\k\ne 0$, event horizon in the maximally extended spacetime is a branch of a \emph{bifurcate Killing horizon}. Again, these results can be derived without using any field equations.

Since null geodesics and Killing vector fields are purely metric notions, they can be directly transferred to PG. Thus, the form of surface gravity \eq{2.5} and the associated zeroth law of black mechanics, are not specific for GR, they are also valid \emph{in the framework of PG.} This aspect of surface gravity was clearly recognized by Chen and Nester \cite{x15}.

The calculation of $\k$ from \eq{2.5} should be done in coordinates that are well defined on the outer horizon. In particular, the metric of static and spherically symmetric black holes in the ingoing Edington-Finkelstein coordinates reads
\be
ds^2=N^2dv^2-2dv\,dr-r^2 d\Om^2\, ,\qquad N=N(r)\, ,
\ee
the Killing vector is $\xi=\pd_v$, and the definition \eq{2.5} of surface gravity takes the form
\be
\pd_r N^2=2\k\, .                                                   \lab{2.7}
\ee

\section{Hamiltonian analysis of PG} 
\setcounter{equation}{0}

In GR, classical black hole entropy can be interpreted as the Noether charge on horizon \cite{x23,x26}. In order to examine this idea in the framework of PG, we find it natural to rely on the Hamiltonian approach, where the canonical charge is derived from the improved form of the gauge generator.

\subsection{First order Lagrangian}

In PG, the conserved charges (energy-momentum and angular momentum) are determined as the values of the (improved) canonical generators of spacetime symmetries, associated to suitable asymptotic conditions \cite{x3,x4}. The canonical procedure is simplified by transforming the quadratic Lagrangian \eq{2.3} into the ``first order" form \cite{x12}
\be
L_G=T^i\t_i+\frac{1}{2}R^{ij}\r_{ij}-V(b^i,\t_i,\r_{ij})\, ,       \lab{3.1}
\ee
where both the gravitational potentials $(b^i,\om^{ij})$ and the corresponding ``covariant momenta" $(\t_i,\r_{ij})$, are \emph{independent} dynamical variables. The potential $V$ is a quadratic function of $(\t_i,\r_{ij})$ which ensures the on-shell relations $\t_i=H_i$ and $\r_{ij}=H_{ij}$, see Appendix \ref{appB}.

In the tensor formalism, the Lagrangian density reads
\be
\tcL_G=-\frac{1}{4}\ve^{\m\n\l\r}
    \left(T^i{}_{\m\n}\t_{i\l\r}+\frac{1}{2}R^{ij}{}_{\m\n}\r_{ij\l\r}\right)
                               -\tcV(b,\t,\r)\, .
\ee
The gravitational field equations (in vacuum) are obtained by varying $\tcL_G$ with respect to the independent dynamical variables $b^i{_\m},\om^{ij}{_\m},\t^i{}_{\m\n}$ and $\r^{ij}{}_{\m\n}$:
\bsubeq\lab{3.3}
\bea
&&\nab_\m\dual{\t}_i{}^{\m\n}-\frac{\pd\tcV}{\pd b^i{_\n}}=0\,, \lab{3.3a}\\[2pt]
&&2b_{[j\m}\dual{\t}_{i]}{}^{\m\n}+\nab_\m\r_{ij}{}^{\m\n}=0\,, \lab{3.3b}\\
&&-\dual{T}^{i\m\n}-\frac{\pd\tcV}{\pd\t_{i\m\n}}=0\, ,    \lab{3.3c}\\[2pt]
&&-\dual{R}^{ij\m\n}  -\frac{\pd\tcV}{\pd\r_{ij\m\n}}=0\,,      \lab{3.3d}
\eea
\esubeq
where we use the notation
$\dual{\t}_i{}^{\m\n}:=\frac{1}{2}\ve^{\m\n\l\r}\t_{i\l\r}$, and similarly
for $\dual{\r}_{ij}{}^{\m\n}$, $\dual{T}^{i\m\n}$ and $\dual{R}^{ij\m\n}$.

\subsection{Primary constraints and Hamiltonians}

Since the Lagrangian $\tcL_G$ describes a gauge invariant dynamical system, transition to the Hamiltonian formalism is characterized by the existence of constraints \cite{x3,x27}. Starting with the field variables $\vphi^A=(b^i{_\m},\om^{ij}{_\m},\t^i{}_{\m\n},\r^{ij}{}_{\m\n})$ and the corresponding canonical momenta
$\pi_A=(\pi_i{^\m},\pi_{ij}{^\m},P_i{}^{\m\n},P_{ij}{}^{\m\n})$, one obtains the following primary constraints:
\bea
&&\phi_i{^0}:=\pi_i{^0}\approx 0\, ,\hspace{31pt}
  \phi_i{^\a}:=\pi_i{^\a}+\dual{\t}_i{}^{0\a}\approx 0\, ,              \nn\\
&&\phi_{ij}{^0}:=\pi_{ij}{^0}\approx 0\, ,\qquad
  \phi_{ij}{^\a}:=\pi_{ij}{^\a}+\frac{1}{2}\,\dual{\r}_{ij}{}^{0\a}\approx 0\,,\nn\\
&&P_i{}^{\m\n}\approx 0\, ,\hspace{59pt} P_{ij}{}^{\m\n}\approx 0\, .
\eea
The canonical Hamiltonian is found to have the form
\bsubeq\lab{3.5}
\bea
&&H_c'=H_c+\pd_\a D^\a\, ,                                         \nn\\
&&H_c:=b^i{_0}\cH_i+\frac{1}{2}\om^{ij}{_0}\cH_{ij}+\t_{i0\a}\dual{T}^{i0\a}
                +\frac{1}{2}\r_{ij0\a}\dual{R}^{ij0\a}+\tcV\,,
\eea
where
\bea
&&\cH_i:=\nab_\a\dual{\t}_i{}^{0\a}\, ,                                  \nn\\
&&\cH_{ij}:=2b_{[j\a}\dual{\t}_{i]}{}^{0\a}+\nab_\a\dual{\r}_{ij}{}^{0\a}\,,\nn\\
&&D^\a:=-b^i{_0}\dual{\t}_i{}^{0\a}
                -\frac{1}{2}\om^{ij}{_0}\dual{\r}_{ij}{}^{0\a}\, . \lab{3.5b}
\eea
\esubeq
Time evolution of dynamical variables is determined by the total Hamiltonian
\be
H_T:=H_c+u^i{_\m}\phi_i{^\m}+\frac{1}{2}u^{ij}{}_\m\phi_{ij}{^\m}
       +\frac{1}{2}v^i{}_{\m\n}P_i{}^{\m\n} +\frac{1}{4}v^{ij}{}_{\m\n}P_{ij}{}^{\m\n}\,,
\ee
where $u$'s and $v$'s are canonical multipliers.

\subsection{Consistency conditions}

The dynamical evolution of the primary constraints $X_A=-(\pi_i{^0},\pi_{ij}{^0},P_i{}^{0\a},P_{ij}{}^{0\a})$ is defined by the consistency conditions $\dot X_A=\{X_A,H_T\}\approx 0$. They produce the secondary constraints
\bea
&&\hcH_i:=\cH_i+\frac{\pd\tcV}{\pd b^i{_0}}\approx 0\,,               \nn\\
&&\hcH_{ij}:=\cH_{ij}\approx 0\, ,                                    \nn\\
&&\hcT^{i0\a}:=\dual{T}^{i0\a}+\frac{\pd\tcV}{\pd\t_{i0\a}}\approx 0\,,  \nn\\
&&\hcR^{ij0\a}:=\dual{R}^{ij0\a}
                +\frac{\pd\tcV}{\pd\r_{ij0\a}}\approx 0\,,          \lab{3.7}
\eea
which correspond to certain components of the field equations \eq{3.3}.

The remaining primary constraints $Y_A=(\phi_i{^\a},\phi_{ij}{^\a},P_i{}^{\a\b},P_{ij}{}^{\a\b})$ are second class, as follows from their Poisson brackets
\be
\{\phi_i{^\g},P_j{}^{\a\b}\}=\eta_{ij}\ve^{0\g\a\b}\d\, ,\qquad
\{\phi_{ij}{}^\a,P_{kl}{}^{\b\g}\}=\eta_{[ik}\eta_{j]l}\ve^{0\a\b\g}\d\,.
\ee
Their consistency conditions determine the canonical multipliers $(u^i{_\a},u^{ij}{_\a},v^i{}_{\a\b},v^{ij}{}_{\a\b})$.~However, we find it more convenient to construct the corresponding Dirac brackets and use them in the consistency procedure on the reduced phase space $\bR$, defined by $Y_A=0$. The only nontrivial Dirac brackets (different from their Poisson counterparts) are
\be
\{b^i{_\a},\t_{j\b\g}\}^*=\d^i_j\ve_{0\a\b\g}\, ,\qquad
\{\om^{ij}{}_{\a},\r_{kl\b\g}\}^*=\d^{[i}_k\d^{j]}_l\ve_{0\a\b\g}\,.
\ee
The form of the total Hamiltonian on $\bR$ is simplified:
\be
H_T=H_c+u^i{_0}\pi_i{^0}+\frac{1}{2}u^{ij}{}_0\pi_{ij}{^0}
       +v^i{}_{0\b}P_i{}^{0\b}
       +\frac{1}{2}v^{ij}{}_{0\b}P_{ij}{}^{0\b}\, .                 \lab{3.10}
\ee
Note that $H_c$ can be  expressed also in terms of the secondary constraints \eq{3.7}
\be
H_c=b^i{_0}\hcH_i+\frac{1}{2}\om^{ij}{_0}\cH_{ij}+\t_{i0\a}\hcT^{i0\a}
      +\frac{1}{2}\r_{ij0\a}\hcR^{ij0\a}.                           \lab{3.11}
\ee
which follows from the $\m=0$ component of the identity
\be
\d^0_\m\tcV=b^i{_\m}\frac{\pd\tcV}{\pd b^i{_0}}
           +\t_{i\m\a}\frac{\pd\tcV}{\pd\t_{i0\a}}
           +\frac{1}{2}\r_{ij\m\a}\frac{\pd\tcV}{\pd\r_{ij0\a}}\,. \lab{3.12}
\ee

Continuing this procedure, one would have to find the consistency conditions of the secondary constraints \eq{3.7}, and so on. However, since our main goal is to construct the gauge generator, we shall follow a simpler approach, described in Appendix \ref{appC}. Once the gauge generator is found, one can construct its improved form \cite{x4,x11}, which defines not only the standard canonical charge, but also black hole entropy.

\section{Entropy and torsion} 
\setcounter{equation}{0}

The Hamiltonian formulation of gravity is based on the existence of a family of spacelike hypersurfaces $\S$, labeled by the time parameter $t$. Each $\S$ is bounded by a closed 2-surface at spatial infinity, which is used to define the \emph{asymptotic charge}. When $\S$ is a black hole manifold, it also possesses an ``interior" boundary, the horizon, which serves to define \emph{black hole entropy} \cite{x23,x26}.

\subsection{Canonical charge as a surface term at infinity}

In PG, conserved charges (energy-momentum and angular momentum) are closely related to the canonical gauge generator $G$, the general form of which is given in Eq. \eq{C.1}. Since $G$ acts on dynamical variables via the Poisson (or Dirac) bracket operation, it should have well-defined functional derivatives on the phase space. In general, $G$ does not satisfy this requirement, but the problem can be solved by adding a suitable surface term $\G_\infty$, located at the boundary of $\S$ at infinity, such that $\tG=G+\G_\infty$ is well defined. The value of $\G_\infty$ is exactly the canonical charge of the system \cite{x3,x4}.

Before continuing with the construction of $\G_\infty$, one should clarify the importance of asymptotic conditions. Local symmetries of PG are characterized by a Killing-Lorentz pair $(\xi^\m,\th^{ij})$, where $\xi^\m$ is a Killing vector that corresponds to local spacetime translations, and $\th^{ij}$ describes local Lorentz rotations. Any particular solution of PG is characterized by a set of asymptotic conditions for basic dynamical variables. Demanding that local Poincar\'e transformations preserve these conditions, one obtains certain restrictions on the Killing-Lorentz parameters. The restricted parameters define the asymptotic symmetry, which is essential for the existence and type of conserved charges.

As we mentioned above, the boundary term $\G_\infty$ is introduced so as to ensure the differentiability of the gauge generator $G=G[\xi,\th]$. To see how this happens, consider the variation of the gauge generator \eq{C.1}
\bea
&&\d G=\int_\S d^3 x (\d G_1+\d G_2)\, ,                                 \nn\\
&&\d G_1=\xi^\m\Big[b^i{_\m}\d\hcH_i+\frac{1}{2}\om^{ij}{_\m}\d\cH_{ij}
  +\t_{i\m\a}\d\hcT^{i0\a}+\frac{1}{2}\r_{ij\m\a}\d\hcR^{ij0\a}\Big]+R\,,\nn\\
&&\d G_2=\frac{1}{2}\th^{ij}\d\cH_{ij}+R\, ,                      \lab{4.1}
\eea
where $\d$ is the variation over the set of asymptotic states, and $R$ denotes regular (differentiable) terms. Next, using the relations \eq{3.7} one obtains
\bea
&&\d G_1=\frac{1}{2}\ve^{0\a\b\g}\xi^\m\Big[b^i{_\m}\nab_\a\d\t_{i\b\g}
            +\frac{1}{2}\om^{ij}{_\m}\nab_\a\d\r_{ij\b\g}
            +2\t_{i\m\g}\nab_\a\d b^i{_\b}
            +\r_{ij\m\g}\nab_\a\d\om^{ij}{_\b} \Big]+R\,,          \nn\\
&&\d G_2=\frac{1}{2}\ve^{0\a\b\g}
            \Big[\frac{1}{2}\th^{ij}\nab_\a\d\r_{ij\b\g}\Big]+R\,. \nn
\eea
To get rid of the unwanted $\d\pd_\m\vphi$ terms which spoil the differentiability of $G$, one can perform a partial integration, which yields
\bea
&&\d G_1=\frac{1}{2}\ve^{0\a\b\g}\pd_\a\Big\{\xi^\m\Big[b^i{_\m}\d\t_{i\b\g}
            +\frac{1}{2}\om^{ij}{_\m}\d\r_{ij\b\g}
            +2\t_{i\m\g}\d b^i{_\b}
            +\r_{ij\m\g}\d\om^{ij}{_\b} \Big]\Big\}+R\, ,          \nn\\
&&\d G_2=\frac{1}{2}\ve^{0\a\b\g}\pd_\a
                    \Big[\frac{1}{2}\th^{ij}\d\r_{ij\b\g}\Big]\,.  \nn
\eea
Integration with the help of Stocke's theorem transforms the above expressions to boundary integrals. Going over to the notation of differential forms ($\ve^{0\a\b\g}d^3x\to-dx^\a dx^\b dx^\g$), the result takes the form
\bsubeq\lab{4.2}
\bea
\d G&=&-\d\G_\infty+R\, ,                                       \lab{4.2a}\\
\d\G_\infty&:=&\oint_{S_\infty}\d B\, ,                         \lab{4.2b}\\
\d B &:=&(\xi\inn b^i)\d H_i+\d b^i(\xi\inn H_i)
                +\frac{1}{2}(\xi\inn\om^{ij})\d H_{ij}
                +\frac{1}{2}\d\om^{ij}(\xi\inn H_{ij})            \nn\\
      &&+\frac{1}{2}\th^{ij}\d H_{ij}\, ,                       \lab{4.2c}
\eea
\esubeq
where $S_\infty$ is the boundary of $\S$ at infinity.

If the adopted asymptotic conditions ensure $\G_\infty$ to be a \emph{finite} solution of the variational equation \eq{4.2}, the improved gauge generator  \be
\tG:=G+\G_\infty
\ee
has well-defined functional derivatives. Then, since $G\approx 0$, the value of $\tG$ is effectively given by the value of $\G_\infty$, which represents the canonical \emph{charge at infinity}. In the particular case $\xi=\pd_t,\th^{ij}=0$, the canonical charge is energy, $E=\G_\infty[\pd_t,\th^{ij}=0]$.

\bitem
\item[(a1)] In the above variational equations, the variation of $\G_\infty$ is defined over a suitable set of asymptotic states, keeping the background configuration fixed.
\eitem
In practical applications, this rule might be sensitive to the choice of coordinates. Nester and co-workers \cite{x15,x16} succeeded to explicitly construct a set of finite expressions $\G_\infty$, which have been tested on a rather wide class of asymptotic conditions. Although their approach yields highly reliable expressions for the conserved charges, we shall continue using the variational approach \eq{4.2}, as it can be naturally extended to a new definition of black hole entropy.

\subsection{Entropy as the canonical charge on horizon}

In order to interpret black hole entropy as the canonical charge on horizon, we assume that the boundary of $\S$ has two components, one at spatial infinity and the other at horizon, $\pd\S=S_\infty\cup S_H$. In an early application of this idea to PG, Nester and co-workers \cite{x15} introduced entropy from a kind of  Hamiltonian conservation law, but with  limited success. The same idea was used later in \cite{x19} to obtain entropy of the BTZ black hole.

Given the spatial hypersurface $\S$ with two boundaries, the condition of
differentiability of the canonical generator $G$ includes two boundary terms, the integrals of $\d B=\d B(\xi,\th)$ over $S_\infty$ and $S_H$:
\be
\d G=-\oint_{S_\infty}\d B+\oint_{S_H}\d B+R\, .               \lab{4.3}
\ee
The sign change in the second term is due to a different orientation of $S_H$. Here, as we already know, the first term represents the asymptotic canonical charge,
\be
\d\G_\infty=\int_{S_\infty}\d B\,,                             \lab{4.4}
\ee
whereas the second one defines entropy $S$ as the canonical \emph{charge on horizon},
\be
\d\G_H:=\oint_{S_H}\d B\, .                                    \lab{4.6}
\ee
\bitem
\item[(a2)] The variation of $\G_H$ is performed by varying the parameters of a solution, but keeping surface gravity constant, in accordance with the zeroth law.
\eitem

Explicit form of entropy depends on two factors:
dynamical and geometric properties of a theory (Riemannian GR, Riemann-Cartan PG, teleparallel theory, etc.), and specific structure of the black hole (static, stationary, etc.).  For stationary black holes in GR, the entropy formula \eq{4.6} takes the well-known form
\be
\d\G_H=T\d S\, ,
\ee
where $T=\k/2\pi$ represents the temperature and $S=\pi r_+^2$ is black hole entropy.

Returning to Eq. \eq{4.3}, one concludes that the gauge generator $G$ is regular if and only if the sum of two boundary terms vanishes,
\be
\d\G_\infty-\d\G_H=0\, ,                                            \lab{4.8}
\ee
which is nothing but the \emph{first law} of black hole thermodynamics. Thus, the validity of the first law directly follows from the regularity of the original gauge generator $G$.

In the framework of PG, the conserved charge is a well-established concept which has been calculated for a number of exact solutions \cite{x8,x13,x14}. In contrast to that, much less is known about black hole entropy. In the next two sections, we will test our definition of black hole entropy \eq{4.6} and the associated first law \eq{4.8}, on three illustrative examples from the family of Schwarzschild-AdS solutions, where AdS stands for anti-de Sitter.

\section{Riemannian Schwarzschild-AdS solution} 
\setcounter{equation}{0}

All exact solutions of GR are also solutions of PG, except for some degenerate cases  \cite{x24}. However, certain properties of a solution may change when we go from GR to a new dynamical environment of PG. As the first test of our results for black hole entropy and the first law, we discuss the case of the Riemannian Schwarzschild-AdS black hole in PG, based on the standard Schwarzschild-like coordinates $x^\m=(t,r,\th,\vphi)$.

\subsection{Geometry}

Riemannian geometry of the Schwarzschild-AdS spacetime is defined by the metric \be
ds^2=N^2dt^2-\frac{dr^2}{N^2}-r^2 (d\th^2+\sin^2\th d\vphi^2)\, , \qquad
N^2:=1-\frac{2m}{r}+\l r^2\, ,                                    \lab{5.1}
\ee
and $\l>0$. The zeros of $N^2$ determine the event horizon
\bsubeq
\be
\l r^3+r-2m=0\, .                                                 \lab{5.2a}
\ee
Since the discriminant of this cubic equation is negative, $\D=-4\l-108\l^2m^2<0$, the equation has just one real root $r_+$. The relation
\be
2m=r_+(\l r_+^2+1)\, .                                            \lab{5.2b}
\ee
\esubeq
implies that $r_+$ is positive iff $m>0$, and $N^2$ is positive in the region $r>r_+$, where the Schwarzschild-like coordinates are well defined.
The surface gravity and black hole temperature are obtained from Eq. \eq{2.7} as
\be
\k=\frac{1}{2r_+}(3\l r_+^2+1)\, ,\qquad T=\frac{\k}{2\pi}\,.     \lab{5.3}
\ee

The orthonormal tetrad is chosen in the form
\be
b^0=Ndt\, ,\qquad b^1=\frac{dr}{N}\, ,\qquad
b^2=rd\th\, ,\qquad b^3=r\sin\th d\vphi\, ,                       \lab{5.4}
\ee
and the horizon area is determined by
\be
A=\int_{S_H}b^2b^3=4\pi r_+^2\, .
\ee
The Riemannian connection reads
\be
\om^{01}=-N' b^0\, ,\qquad \om^{1c}=\frac{N}{r}b^c\, ,\qquad
\om^{23}=\frac{\cos\th}{r\sin\th}b^3\, ,
\ee
and the corresponding curvature 2-form $R^{ij}$ has two nonvanishing irreducible pieces
\bea
&&{}^{(6)}R^{ij}=\l b^ib^j\, ,                                     \nn\\
&&{}^{(1)}R^{01}=-\frac{2m}{r^3}b^0b^1\, ,\qquad
  {}^{(1)}R^{23}=-\frac{2m}{r^3}b^2b^3\, ,\qquad
  {}^{(1)}R^{Ac}=\frac{m}{r^3}b^Ab^c\,.
\eea
The covariant momenta are $H_i=0$ and
\bea
&&H_{01}=-2b^2b^3\Big(a_0+2b_1\frac{m}{r^3}-b_6\l\Big)\,,\qquad
  H_{23}=-2b^0b^1\Big(a_0+2b_1\frac{m}{r^3}-b_6\l\Big)\, ,           \nn\\
&&H_{Ac}=-\ve_{Acmn}b^mb^n\Big(a_0-b_1\frac{m}{r^3}-b_6\l\Big)\,. \lab{5.8}
\eea
In these formulas, we use a convenient 2+2 splitting $i=(A,c)$, where $A=0,1$ and $c=2,3$.

Based on the PG field equations \eq{2.1}, one can show that the Riemannian Schwarzschild-AdS spacetime is an exact solution of PG, provided that
\be
3a_0\l+\L=0\, .
\ee

\subsection{Entropy and the first law}

Energy of the Riemannian Schwarzschild-AdS solution in PG can be calculated from the variational formula \eq{4.2c} for $\xi=\pd_t$ and $\th^{ij}=0$. The result is given by (Appendix \ref{appD})
\be
E=16\pi A_0 m, \qquad A_0:=a_0+\l(b_1-b_6)\, .
\ee
For energy at horizon, the variational formula \eq{4.6} defines entropy as follows:
\bsubeq
\bea
&&\d \G_H=\oint_{S_H}\om^{01}{_t}\d H_{01}=8\k A_0\,\d(\pi r_+^2)\,,     \\
\Ra&& \d \G_H=T\d S\, ,\qquad  S=16\pi A_0(\pi r_+^2)\, .
\eea
\esubeq

Since Eqs. \eq{5.2b} and \eq{5.3} imply $2\d m=\k\d r_+^2$, we have $\d E=\d\G_H$, which confirms the validity of the first law
\be
\d E=T\d S\, .                                                  \lab{5.12}
\ee
The presence of the multiplicative factor $A_0\ne a_0$ shows that
entropy of the Schwarzschild-AdS black hole in PG, as well as the first law, agrees with the corresponding result for diffeomorphism invariant Riemannian theories, see Wald \cite{x23} and Jacobson \cite{x26}.

\subsection{Reduction to GR}

The GR limit is recovered for $b_1=b_6=0$, $A_0=a_0$ and $16\pi a_0=1$:
\be
E=m\, ,\qquad S=\pi r_+^2\, .
\ee

\section{Schwarzschild-AdS solutions with torsion} 
\setcounter{equation}{0}

As is well known from our experience with GR, exact solutions have an essential role in revealing hidden aspects of the gravitational dynamics.
The dynamics of the (parity invariant) PG is defined by a Lagrangian with ten coupling constants, which makes the search for exact solutions a rather complicated task. Despite that, many of the known GR solutions have been successfully generalized to the corresponding solutions with torsion; for more details, see the reader \cite{x3}. In this section, we shall examine two spherically symmetric solutions with torsion.

\subsection{Baekler solution} 

One of the first spherically symmetric solutions of PG has been constructed by Baekler \cite{x28}. We shall use this solution to verify our approach to entropy in the presence of torsion.

\subsubsection*{Formulation of the model}

The metric of the Baekler solution is of the Schwarzschild-AdS form, with the tetrad field given as in Eq. \eq{5.4}. The ansatz for torsion is assumed to be $O(3)$ invariant (rotations and reflections) \cite{x29}. More specifically,
\bsubeq\lab{6.1}
\bea
&& T^0=T^1=f b^0b^1\, ,\qquad T^c=-f (b^0-b^1)b^c\, ,
\eea
where $f$ is a function of $r$,
\be
f:=-\frac{m}{r^2N}\, .
\ee
\esubeq
The third irreducible component of $T^i$ vanishes, ${}^{(3)}T^i=0$.

Having adopted the anzatz for torsion, one can calculate the Riemann-Cartan connection
\bea
&&\om^{01}=-(N'+f)b^0+fb^1\, ,\qquad \om^{0c}=-f b^c\, ,         \nn\\
&&\om^{1c}=\left(\frac{N}{r}-f\right)b^c\, , \qquad
  \om^{23}=\frac{\cos\th}{r\sin\th}b^3\, ,                          \lab{6.2}
\eea
whereupon the curvature 2-form turns out to have only two nonvanishing irreducible parts,
\be
\ir{6}R^{ij}=\l b^ib^j\, ,\qquad
\ir{4}R^{Ac}=\frac{\l m}{rN^2}(b^0-b^1)b^c\, .                     \nn
\ee

Dynamics is determined by a two-parameter PG Lagrangian
\bea
L_G=a_1 T^i\hd({}^{(1)}T_i-2\,{}^{(2)}T_i+{}^{(3)}T_i)
    +\frac{1}{2}b_1R^{ij}\hd R_{ij}\,,                             \lab{6.3}
\eea
proposed by von der Heyde \cite{x30}. The field equations \eq{2.1} produce the following restriction on the Lagrangian parameters:
\be
2\l b_1=-a_1\, .
\ee

\subsubsection*{Entropy and the first law}

The form of surface gravity is the same as in Eq. \eq{5.3}.
Explicit expressions for the covariant momenta $H_{ij}=2b_1\hd R_{ij}$ and $H_i=2a_1\hd(\ir{1}T_i-2\ir{2}T_i)$ read:
\bea
&&H_{01}=-a_1b^2b^3\, ,\qquad H_{23}=-a_1b^0b^1\, ,                \nn\\
&&H_{02}=a_1b^1b^3-a_1\frac{m}{rN^2}(b^0-b^1)b^3\, ,               \nn\\
&&H_{03}=-a_1b^1b^2+a_1\frac{m}{rN^2}(b^0-b^1)b^2\, ,              \nn\\
&&H_{12}=-a_1b^0b^3+a_1\frac{m}{rN^2}(b^0-b^1)b^3\, ,              \nn\\
&&H_{13}=a_1b^0b^2-a_1\frac{m}{rN^2}(b^0-b^1)b^2\, ,               \nn\\
&&H_0=-H_1=4a_1\frac{m}{r^2N}b^2b^3\,.
\eea
Energy of the solution is proportional to $m$ (Appendix \ref{appD}):
\be
E=16\pi a_1 m\, .                                                \lab{6.6}
\ee
Entropy is calculated from the variational equation \eq{4.6} (integration implicitly understood):
\bea
&&b^i{_t}\d H_i=-4\big[N\d(fr^2)\big]_{r_+}\cdot 4\pi a_1\,,                       \nn\\
&&\frac{1}{2}\om^{ij}{_t}\d H_{ij}
  =\om^{01}{_t}\d H_{01}=(\k+\ul{Nf}_\times)\d r_+^2\cdot 4\pi a_1 \nn\\
&&\frac{1}{2}\d\om^{ij}H_{ijt}
  =\Big[-2fr^2\d N+2N\d(fr^2)-\ul{Nf}_\times\d r^2\Big]_{r_+}\cdot 4\pi a_1\,.
                                                                  \lab{6.7}
\eea
Summing up these terms, one can identify black hole entropy:
\be
\d\G_H=8\pi a_1\k\d r_+^2=T\d S\, ,\qquad
           S:=16\pi a_1\d(\pi r_+^2)\, .                          \lab{6.8}
\ee
Then, the identity  $2\d m-\k\d r_+^2=0$ implies the validity of the first law:
\be
\d\G_\infty=\d\G_H\quad\Lra\quad \d E=T\,\d S\,.
\ee

With the normalization $16\pi a_1=1$, both energy and entropy reduce to the corresponding GR expressions. However, since the torsion sector gives a nontrivial contribution to entropy, see the first line in \eq{6.7}, \emph{dynamical content} of the result is quite different. The approach of Refs. \cite{x15} yields a different expression for entropy, which is incompatible with the first law.

\subsection{Schwarzschild-AdS solution in teleparallel gravity} 

Teleparallel gravity (TG) is a subcase of PG, defined by the vanishing Riemann-Cartan curvature, $R^{ij}=0$. Choosing the related spin connection to vanish, $\om^{ij}=0$, the tetrad field remains the only dynamical variable, and torsion takes the form $T^i=d b^i$. The general (parity invariant) TG Lagrangian has the form
\bsubeq
\be
L_T:=a_0T^i\,\hd\left(a_1\ir{1}T_i+a_2\ir{2}T_i+a_3\ir{3}T_i\right)\,.
\ee
In physical considerations, a special role is played by a special \emph{one-parameter family} of TG Lagrangians \cite{x3,x4}, defined by a single parameter $\g$ as
\be
a_1=1\,,\qquad a_2=-2\,,\qquad a_3=-1/2+\g\,.
\ee
\esubeq
This family represents a viable gravitational theory for macroscopic matter, empirically indistinguishable from GR.

Every spherically symmetric solution of GR is also a solution of the one-parameter TG. In particular, this is true for the Schwarzschild-AdS spacetime. Since $\ir{3}T_i=0$, the covariant momentum $H^i$ does not depend on $\g$:
\bea
&&H^0=\frac{2a_0}{r\sin(\th)}\Big[\cos(\th)b^1b^3-2N\sin(\th)b^2b^3\Big]\,,\nn\\
&&H^1=\frac{2a_0\cos(\th)}{r\sin(\th)}b^0b^3\, ,                    \nn\\
&&H^2=-\frac{2a_0}{r}(rN'+N)b^0b^3\, ,                              \nn\\
&& H^3=\frac{2a_0}{r}(rN'+N)b^0b^2\,.
\eea

The energy of the Schwarzschild-AdS solution in TG is (Appendix \ref{appD}):
\be
E=m\, .                                                            \lab{6.12}
\ee
When the only nontrivial covariant momentum is $H_i$, the entropy formula proposed in Refs. \cite{x15} does not work. Our approach to entropy yields (integration implicit)
\bsubeq\lab{6.13}
\bea
&&b^i{_t}\d H_i=\big[N\d H_0\big]_{r_+}
                        =-16\pi a_0\big[N\d(Nr)\big]_{r_+}=0\, ,    \nn\\
&&b^i\d H_{it}=\big[b^2\d H_{2t}+b^3\d H_{3t}\big]_{r_+}
              =8\pi a_0\cdot\k\d(r_+^2)\, ,
\eea
where we used $NN'=\k$ and $[N\d N]_{r_+}=0$. Thus, with $16\pi a_0=1$, one obtains
\be
\d\G_H=T\d S\, ,\qquad S=\pi r_+^2\, .
\ee
\esubeq
The identity $2\d m=\k\d r_+^2$ confirms the validity of the first law
\be
\d E=T\d S\, .
\ee

\section{Concluding remarks}
\setcounter{equation}{0}

In the present paper, we investigated the notion of entropy in the  general (parity preserving) four-dimensional PG. Our approach was based on the idea that black hole entropy can be interpreted as the conserved charge on horizon; see Wald \cite{x23}.

To examine this idea in the framework of the Hamiltonian formalism, we constructed the canonical generator $G$ of gauge symmetries as an integral over the spatial section $\S$ of spacetime. Since $G$ acts on the dynamical variables via the Poisson bracket operation, it has to be a regular (differentiable) functional on the phase space. An analysis along the lines proposed by Regge and Teitelboim \cite{x11} shows that regularity can be ensured by adding to $G$ a suitable surface term $\G_\infty$ defined on the boundary of $\S$ at infinity. Combined with suitable boundary conditions, the form of $\G_\infty$ is determined by the variational equation \eq{4.2} and its value defines the asymptotic charge of a PG solution.

For a black hole solution, $\S$ has two boundaries, one at infinity and the other at horizon, and the condition of regularity of $G$ includes two boundary terms, $\G_\infty$ and $\G_H$. The new boundary term $\G_H$, determined by the variational equation \eq{4.6}, defines entropy as the canonical charge on horizon. Moreover, the regularity of $G$ is expressed by the condition $\G_\infty-\G_H=0$, which is just the first law of black hole thermodynamics.

We tested our results on three vacuum solutions of the Schwarzschild-AdS type. Treating the Riemannian Schwarzschild-AdS geometry as a solution of PG, we found that both energy and entropy differ from the GR expressions by a multiplicative factor, in agreement with earlier results \cite{x23,x26}. In the next step, we analyzed Baekler's solution \cite{x28}, one of the first exact solutions found in PG. The result reveals new dynamical features of PG, the existence of nontrivial contributions to energy and entropy stemming from both the torsion and the curvature sectors. It is astonishing to see how these two sectors interfere to produce the final result which is exactly the same as in GR. In the last example we successfully applied our approach to the teleparallel gravity, where curvature vanishes and entropy is produced solely by torsion, respecting the first law.

An additional test of our approach to black hole entropy can be obtained from the analysis of the Kerr black hole \cite{x31}.

\section*{Acknowledgments}

We would like to thank Friedrich Hehl for his continuing support of our studies of black hole entropy in PG, as well as to James Nester for a critical reading of the manuscript. This work was partially supported by the Serbian Science Foundation under Grant No. 171031. The results are checked using the computer algebra system {\sc Reduce}.

\appendix
\section{Irreducible decomposition of the field strengths}\label{appA}
\setcounter{equation}{0}

We present here formulas for the irreducible decomposition of the Poincar\'e gauge theory field strengths in a 4D Riemann-Cartan spacetime.
The torsion 2-form has three irreducible pieces:
\bea
&&{}^{(2)}T^i=\frac{1}{3}b^i\wedge(h_m\inn T^m)\, ,             \nn\\
&&{}^{(3)}T^i=\frac{1}{3}h^i\inn(T^m\wedge b_m)\, ,             \nn\\
&&{}^{(1)}T^i=T^i-{}^{(2)}T^i-{}^{(3)}T^i\, .
\eea
The RC curvature 2-form can be decomposed into six irreducible pieces:
\bsubeq
\be
\ba{ll}
{}^{(2)}R^{ij}={}^*(b^{[i}\wedge\Psi^{j]})\, ,
           &{}^{(4)}R^{ij}=b^{[i}\wedge\Phi^{j]}\, ,            \\[3pt]
{}^{(3)}R^{ij}=\dis\frac{1}{12}X\,{}^*(b^i\wedge b^j)\, ,
           &{}^{(6)}R^{ij}=\dis\frac{1}{12}F\,b^i\wedge b^j\, , \\[9pt]
{}^{(5)}R^{ij}=\dis\frac{1}{2}b^{[i}\wedge h^{j]}\inn(b^m\wedge F_m)\,,
   \qquad  &{}^{(1)}R^{ij}=R^{ij}-\sum_{a=2}^6{}^{(a)}R^{ij}\, .\lab{A.2a}
\ea
\ee
where
\bea
&&F^i:=h_m\inn R^{mi}=\ric^i\, , \qquad F:=h_i\inn F^i=R\, ,    \nn\\
&&X^i:={}^*(R^{ik}\wedge b_k)\, ,\qquad X:=h_i\inn X^i\,.       \lab{A.2b}
\eea
and
\bea
&&\Phi_i:=F_i-\frac{1}{4}b_iF-\frac{1}{2}h_i\inn(b^m\wedge F_m)\,,\nn\\
&&\Psi_i:=X_i-\frac{1}{4}b_i X-\frac{1}{2}h_i\inn(b^m\wedge X_m)\, .
\eea
\esubeq

These formulas differ from those in Refs. \cite{x2,x24} in two minor details: The definitions of $F^i$ and $X^i$ are taken with an additional minus sign, but simultaneously, the overall signs of all the irreducible curvature parts are also changed, leaving their final content unchanged.

\section{Explicit form of \mb{V}}\label{appB}
\setcounter{equation}{0}

The torsion part of the potential $V$ can be written in the form \cite{x20}
\be
V_{\t^2}:=\t^i\Big(\m_1\ir{1}\hd\t_i+\m_2\ir{2}\hd\t_i
           +\m_3\ir{3}\hd\t_i\Big)\,,\qquad\m_n=-\frac{1}{4a_n}\, .\lab{B.1}
\ee
Indeed, since the variation of the Lagrangian \eq{3.1} with respect to $\t^i$ yields
\be
T_i=2\big(\m_1\ir{1}\hd\t_i
                      +\m_2\ir{2}\hd\t_i+\m_3\ir{3}\hd\t_i\big)\,, \lab{B.2}
\ee
one obtains
\be
\ir{n}T_i=2\m_n\ir{n}\hd\t_i\quad\Ra\quad
      -2\sum_n a_n\ir{n}T_i=\hd\t_i \quad\Ra\quad H_i=\t_i\, ,     \lab{B.3}
\ee
as expected. Moreover, this result implies
\be
V_{\t^2}=\frac{1}{2}\t^i T_i\quad\Ra\quad
   L_T=T^i\t_i-\frac{1}{2}\t^i T_i=\frac{1}{2}T^i H_i\, .
\ee

Next, note that the duality properties
\be
\ir{1}\hd\t_i=\hd\ir{1}\t_i\, ,\qquad \ir{2}\hd\t_i=\hd\ir{3}\t_i\, ,
\qquad \ir{3}\hd\t_i=\hd\ir{2}\t_i\, ,
\ee
lead to an equivalent form of $V_{\t^2}$:
\be
V_{\t^2}=\t^i\,\hd\Big(\m_1\ir{1}\t_i
                    +\m_3\ir{2}\t_i+\m_2\ir{3}\t_i \Big)\,.
\ee
Transition to the tensor formalism yields
\be
\tcV_{\t^2}=\frac{1}{2}b\t^{imn}
              \big(\m_1\ir{1}\t_{imn}+\m_3\ir{2}\t_{imn}
                                  +\m_2\ir{3}\t_{imn}\big)\,.
\ee

An analogous construction holds also in the curvature sector.

\section{Canonical gauge generator}\label{appC}
\setcounter{equation}{0}

The canonical generator of local Poincar\'e transformations can be constructed
following Castellani's algorithm \cite{x32} based on the knowledge of the Poisson bracket algebra of the first class constraints. However, there is an alternative procedure formulated in Ref. \cite{x10}, according to which
a phase-space functional $G$ is a good gauge generator if it generates the correct gauge transformations of all phase-space variables. Relying on an explicit construction of $G$ in the three-dimensional PG \cite{x20}, we display here its generalization to 4D:
\bsubeq\lab{C.1}
\bea
G[\xi,\th]&=&\int_\S d^3 x\,\big(G_1+G_2\big)\,,                        \nn\\
G_1&=&\dot\xi^\m\Big(b^i{_\m}\pi_i{^0}
  +\frac{1}{2}\om^{ij}{_\m}\pi_{ij}{^0}
  +\t^i{}_{\m\b}P_i{}^{0\b}+\frac{1}{2}\r^{ij}{}_{\m\b}P_{ij}{}^{0\b}\Big)
  +\xi^\m\cP_\m\, ,                                                     \nn\\
G_2&=&\frac{1}{2}\dot\th^{ij}\pi_{ij}{^0}+\frac{1}{2}\th^{ij}\cM_{ij}\,,
\eea
where
\bea
\cP_\m&:=&b^i{_\m}\hcH_i+\frac{1}{2}\om^{ij}{}_\m\cH_{ij}
  +\t^i{}_{\m\b}\hcT_i{}^{0\b}+\frac{1}{2}\r^{ij}{}_{\m\b}\hcR_{ij}{}^{0\b}\nn\\
&&+(\pd_\m b^i{_0})\pi_i{^0}+\frac{1}{2}(\pd_\m\om^{ij}{_0})\pi_{ij}{^0}
        +(\pd_\m\t^i{}_{0\b})P_i{}^{0\b} +\frac{1}{2}(\pd_\m\r^{ij}{}_{0\b})P_{ij}{}^{0\b}               \nn\\
&&-\pd_\b\left(\t^i{}_{0\m}P_i{}^{0\b}
               +\frac{1}{2}\r^{ij}{}_{0\m}P_{ij}{}^{0\b}\right)\, ,     \nn\\
\cM_{ij}&:=&\cH_{ij}+2b_{[i0}\pi_{j]}{^0}+2\om^k{}_{[i0}\pi_{kj]}{^0}
                    +2\t_{[i0\g}P_{j]}{}^{0\g}
                    +2\r^k{}_{[i0\g}P_{kj]}{}^{0\g}\, .
\eea
\esubeq

Now, we wish to show that the action of $G$ on the phase-space variables $\vphi$, defined by $\d_0\vphi:=\{\vphi,G\}^*$, has the correct PG form.
Note that the generator $\cP_0$ is just the total Hamiltonian, $\cP_0=H_T$, as follows from \eq{3.11}; hence, $\{\vphi,\cP_0\}^*=\dot\vphi$.

To find the transformation laws generated by the action of $G=G_1+G_2$, we start with the variables $(b^i{_\a},\om^{ij}{_\a},\t_{i\a\b})$. For $b^i{_\a}$, the result has the form
\bea
&&\d_0 b^i{_\a}=-\th^i{_k}b^k{_\a}+
  (\pd_\a\xi^\m) b^i{_\m}+\xi^\m\pd_\m b^i{_\a}+\xi^\m X^i{}_{\a\m}\,, \nn\\
&&X^i{}_{\a\m}:= T^i{}_{\a\m}
       -\frac{1}{2}\ve_{\a\m\l\r}\frac{\pd\tcV}{\pd\t_{i\l\r}}\,,
\eea
whereas the transformation law for $\om^{ij}{_\a}$ reads
\bea
&&\d_0\om^{ij}{_\a}=\nab_\a\th^{ij}
  +(\pd_\a\xi^\m)\om^{ij}{_\m}+\xi^\m\pd_\m\om^{ij}{_\a}
                        +\xi^\m Y^{ij}{}_{\a\m}\, ,                    \nn\\
&&Y^{ij}{}_{\a\m}:=R^{ij}{}_{\a\m}
  -\frac{1}{2}\ve_{\a\m\l\r}\frac{\pd\tcV}{\pd\r_{ij\l\r}}\,.
\eea
An analogous calculation for $\t_{i\b\g}$ yields
\bea
&&\d_0\t^i{}_{\a\b}=-\th^i{_k}\t^k{}_{\a\b}
  +\pd_{\a}\xi^\m\t^i{}_{\m\b}+\pd_{\b}\xi^\m\t^i{}_{\a\m} +\xi^\m\pd_\m\t^i{}_{\a\b} +\xi^\m Z^i{}_{\a\b\m}\, ,               \nn\\
&&Z^i{}_{\a\b\m}:=\nab_\a\t^i{}_{\b\m}+\nab_{\m}\t^i{}_{\a\b}
  +\nab_\b\t^i{}_{\m\a}+\ve_{\a\b\m\n}\frac{\pd\tcV}{\pd b_{i\n}}\,.
\eea
Since the field equations \eq{3.3} imply $X^i{}_{\a\m},Y^{ij}{}_{\a\m},Z^i{}_{\a\b\m}=0$, one concludes that the above transformations have the correct on-shell form.

Next, the gauge transformations of the pair $(\t_{i0\a},P_i{}^{0\a})$ are given by
\bea
&&\d_0\t_{i0\a}=-\th_i{^k}\t_{k0\a}+(\pd_0\xi^\m)\t_{i\m\a}
         +(\pd_\a\xi^\b)\t_{i0\b}+\xi^\m\pd_\m\t_{i0\a}\,,         \nn\\
&&\d_0 P_{i}{}^{0\a}=-\th_i{^k}P_k{}^{0\a}-(\pd_0\xi^0)P_i{}^{0\a}
         -(\pd_\b\xi^\a)P_i{}^{0\b}+\pd_\m(\xi^\m P_i{}^{0\a})\,.
\eea
These transformations, as well as those found for  $(\r_{ij0\b},\r_{ij\a\b})$ and the remaining momentum variables are of the expected form (all the momenta are tensor densities).

The above results confirm that the expression \eq{C.1} is the correct canonical generator of gauge transformations in PG.

\section{Energy for Schwarzschild-AdS solutions}\label{appD}
\setcounter{equation}{0}

In this appendix, we present the calculations of energy for the Schwarzschild-AdS solutions in vacuum, discussed in the main text. The results are found using the variational formula \eq{4.2} with $\xi=\pd_t$ and $\th^{ij}=0$. The AdS asymptotics allows the variation of the mass parameter $m$, whereas $\l$ remains fixed.

Starting with the asymptotic formulas
\be
N'N=\l r+O_2\, ,\qquad N\d N=-\frac{\d m}{r}\,,
\ee
one finds that the value of energy for the Riemannian Schwarzschild-AdS solution in PG is determined by the following nonvanishing contributions to $\d\G_\infty$ (integration is implicit):
\bsubeq\lab{D.2}
\bea
&&\frac{1}{2}\om^{ij}{_t}\d H_{ij}=\om^{01}{_t}\d H_{01}
          =4\pi N'N\Big(4b_1\frac{\d m}{r}\Big)=16\pi\l b_1\d m\,,   \nn\\
&&\frac{1}{2}\d\om^{ij}H_{ijt}=\d\om^{1c}H_{1c t}
          =-16\pi(\d N) Nr(a_0-\l b_6)=16\pi(a_0-\l b_6)\d m\, .
\eea
Summing up, one obtains
\be
E:=\G_\infty=16\pi \big[a_0+\l(b_1-b_6)\big]m\, .
\ee
\esubeq
Analogously, the calculation for Baekler's solution with torsion yields
\bea
&&b^i{_t}\d H_i=N\d H_0=N\d\Big(4a_1\frac{m}{r^2N}b^2b^3\Big)
                       =16\pi a_1\d m\, ,                             \nn\\
\Ra&&E=16\pi a_1 m\, .                                             \lab{D.3}
\eea
Finally, for the Schwarzschild-AdS solution in the one-parameter TG, the expression for energy takes the GR form
\bea
&&b^i{_t}\d H_i=N\d H_0=-16\pi a_0 rN\d N=\d m\, ,                 \nn\\
\Ra&& E=m\, ,                                                      \lab{D.4}
\eea
where we used $16\pi a_0=1$.

The above results for energy are identical to those obtained using the formulas of Nester and co-worker \cite{x13,x16}.

\end{document}